\documentclass[english,aps]{revtex4-2}
\usepackage[T1]{fontenc}
\usepackage[latin9]{inputenc}
\usepackage{xcolor}
\usepackage{float}
\usepackage{amsmath}
\usepackage{amssymb}
\usepackage{graphicx}

\makeatletter
\usepackage{hyperref}

\makeatother

\usepackage{babel}

\begin{document}
\title{Search for oscillating fundamental constants using a paired detector
and vibrational spectroscopy}
\author{René Oswald, Victor Vogt and Stephan Schiller}
\address{Institut für Experimentalphysik, Heinrich-Heine-Universität Düsseldorf,
40225 Düsseldorf, Germany}
\begin{abstract}
Ultralight dark matter (UDM) may manifest itself through oscillating
fundamental constants of normal matter. These can be experimentally
searched for by implementing two dissimilar oscillators producing
a beat between their frequencies and analyzing the beat-frequency
time series for the presence of any temporal oscillations. Typically,
the time series of such a detector contains contributions from nonstationary
noise. In order to reduce the influence of such noise we propose and
demonstrate paired detectors: two nominally identical detectors whose
signals are synchronously recorded. The cross-spectrum of the two
individual beat time series is then analyzed for UDM signatures. This
approach permits us to suppress spurious signals appearing in uncorrelated
fashion in either detector. We furthermore demonstrate detectors that
are based on a vibrational molecular transition, which are advantageous
due to their larger sensitivity to oscillations of the nuclear masses.
The analysis of 274~hours of data yielded improved bounds for the
coupling constants of UDM to nuclear mass and to electron mass in
the frequency ranges 10--500~Hz and 10--122~kHz, with improvement
factors between 6 and 10. These bounds are currently the strongest.
Similar bounds are obtained for the fine-structure constant. The present
approach may be generalized to large ensembles of detectors.
\end{abstract}
\maketitle

\section{Introduction}

It is today still unknown whether dark matter couples to normal matter
and the mass (or masses) of dark matter (DM) particles are also unknown. In fact,
the allowed mass range is poorly constrained, so far. Shedding light
on these fundamental questions is one of the most important challenges
in fundamental physics today. It is possible that DM particle masses
are well below 1~eV. This so-called ultralight DM (UDM) can be described
by a classical field $\phi(t)$ that, according to a common model
\citep{Foster2018}, oscillates in a nearly monofrequent fashion at
any given location on Earth. The characteristic frequency is the Compton
frequency $f_{\phi}$ of the DM particle. The amplitude of the field
is predicted to vary extremely slowly on the timescale of the oscillation
period \citep{Derevianko2018}.

Furthermore, if UDM couples to normal matter, the coupling may be
in the form of a Lagrangian function for the DM-normal matter interaction
in which $\phi$ couples linearly to the energy densities of normal
matter. Such densities arise from the fields associated with fermionic
matter (electrons, quarks, gluons) and with the electromagnetic field.
The interaction strengths are quantified by a set of coupling constants
${d_{g}}$. As a consequence, the fundamental constants (FCs) become
functions of the dark matter field \citep{Damour2010,Stadnik2015,Antypas2020,Oswald2022,Arvanitaki2016}.\vspace{0bp}
\begin{eqnarray}
m_{X}(\phi) & = & m_{X}\left(1+d_{X}\,\phi\right),\label{eq:v1}\\
\alpha(\phi) & \simeq & \alpha\left(1-d_{\alpha}\,\phi\right),\label{eq:v2}\\
\alpha_{{\rm s}}(\phi) & \simeq & \alpha_{{\rm s}}\left(1-\frac{2d_{g_{{\rm s}}}\beta'(g_{{\rm s}})}{g_{{\rm s}}}\,\phi\right).\label{eq:v3}
\end{eqnarray}
 Here $m_{X}$ are the masses of the fermions, where $X=e,u,d,s$
stand for electron, up-quark, down-quark and strange-quark, respectively.
$\alpha$ is the fine-structure constant, $g_{s}=\sqrt{4\pi\alpha_{s}}$
is the strong-interaction coupling constant, and the function $\beta'(g_{s})$
describes the evolution of the coupling constant with energy.

A consequence of the above is that also the mass of any nucleus $m_{N}$
depends on $\phi$. This dependence involves essentially four coupling
constants, 
\begin{equation}
m_{N}(\phi)\simeq m_{N}\left(1+(Q_{N}\cdot d_{N})\,\phi\right),\label{eq: m_N(phi)}
\end{equation}
with the shorthands $d_{N}=(d_{g_{s}},d_{\hat{m}},d_{\delta m},d_{s})$
and $Q_{N}\simeq0.878\,\times(0.994,0.096,3\times10^{-4},0.049)$.
In this expression, the electromagnetic binding energy of the nucleus
has been neglected, and $\hat{m}=(m_{u}+m_{d})/2$, $\delta m=m_{u}-m_{d}$.

Thus, the FCs inherit the oscillatory character of $\phi$ and may
be (extremely weakly) oscillating ``constants.'' It is the task of
experiments to set bounds to the magnitude of the six coupling constants
$d_{\alpha}$, $d_{m_{e}}$ $d_{N}$. Already, a number of experiments
have been performed to search for such oscillations \citep{Antypas2020,Savalle2021,Filzinger2023,Zhang2023a,Campbell2021,Kennedy2020}.
They have addressed very different ranges of the Compton frequency
$f_{\phi}$, between $10^{-7}$ Hz to 100~MHz. Because most experiments
have been based on cavities and atomic spectroscopy, their sensitivity
to oscillating nuclear masses has been intrinsically small--only
occurring through the reduced-mass effect of atomic transitions (for
a different approach, see \citep{Savalle2021}). The sensitivity has
also been particularly small because typically heavy atoms have been
used. Therefore, recently two experiments have focused on molecular
spectroscopy, where the vibrational energies provide direct access
to the nuclear mass \citep{Oswald2022}. However, because electronic
molecular transitions were interrogated, the contribution from vibrational
energies was partially suppressed.

\begin{figure}[H]
\centering{}\includegraphics[width=1\textwidth]{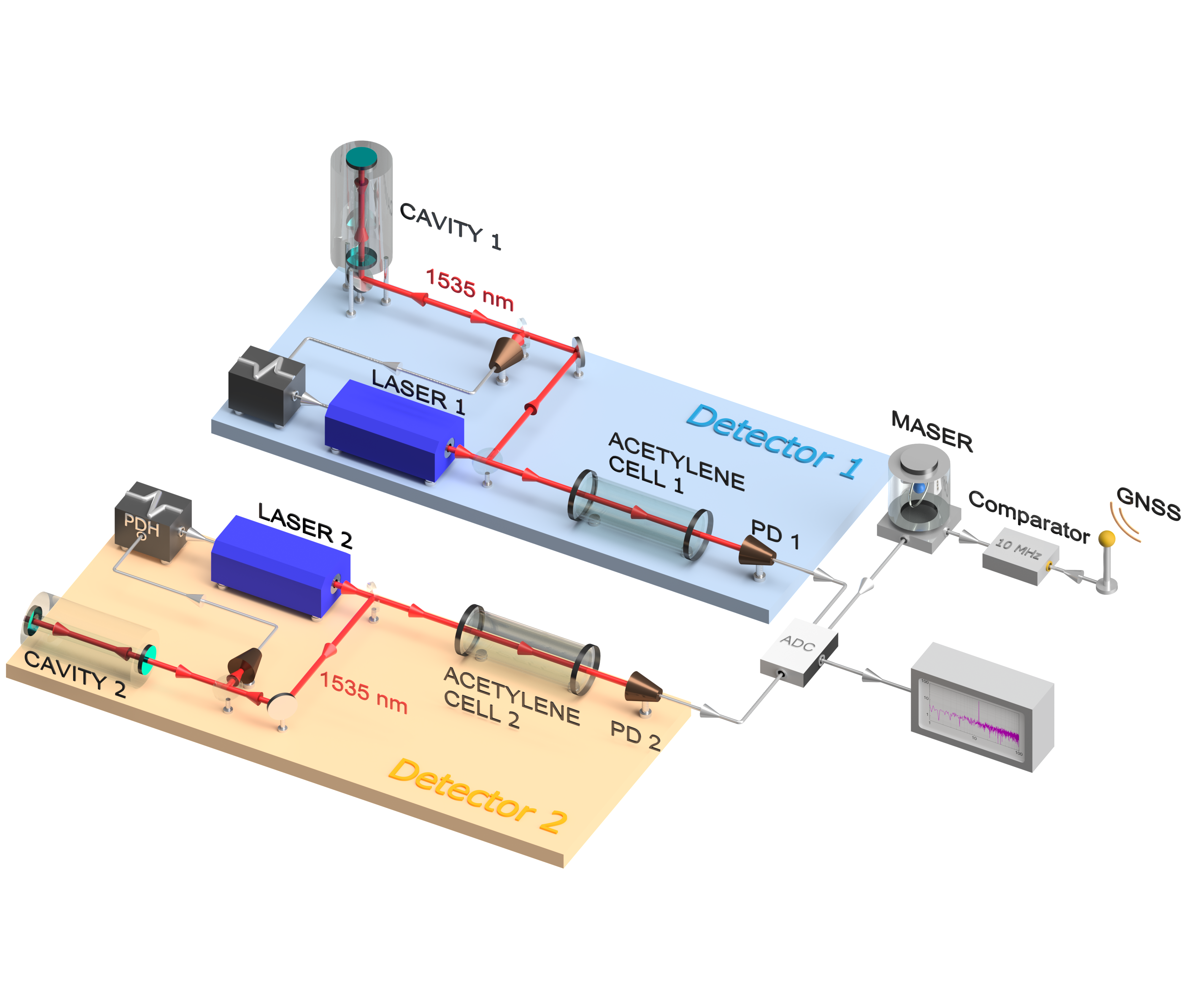}\caption{\label{fig:Principle-of-the-experiment}Concept of the experimental
apparatus. Two nominally identical detectors 1 and 2 generate output
signals on respective photodetectors PD1, PD2 that are synchronously
recorded by an analog-to-digital converter (ADC) and then processed.}
\end{figure}

In this work, we present a new-generation UDM molecular spectroscopy
experiment sketched in Fig.~\ref{fig:Principle-of-the-experiment}.
Its aim is to allow setting stronger bounds for fast oscillations
($f_{\phi}>10\,{\rm Hz}$) of the mentioned coupling constants. Its
innovations are first, the use of a molecular gas in which a vibrational
rather than an electronic transition is probed, leading to a substantial
gain in sensitivity to the nuclear mass and to the electron mass.
Second, two independent (but neighboring) detectors are operated simultaneously,
and their signals are recorded with attention to using a common time
base. Therefore, it is possible to perform a cross-correlation analysis
of the two signals. This allows us to reject spurious signals that occur
in one detector only as well as reducing the impact of uncorrelated
noise present in both detectors. With this demonstration, we take
the first step toward future (concentrated or distributed) arrays
of tens, possibly hundreds or thousands of simultaneously operated
detectors.

In more detail, each detector of a pair is exposed to various disturbances,
e.g. mechanical vibration, acoustic noise, electrical noise. Their
influence on the spectrum of the recorded time series is significant.
In the case of having access only to a single detector, the only approach
for reducing the impact of noise is to repeat data acquisition runs
and then average the spectra over different experiments. This will
reduce the importance of noise uncorrelated between runs, but it will
not remove systematic disturbances, e.g. signals at constant frequency
with slowly varying amplitude. Spurious signals are also difficult
to remove, necessitating the detailed investigation of their spectrum.

A more efficient, but also more costly, approach is to have two
detectors that take data simultaneously. Cross-correlating the data
allows us to discriminate not only against uncorrelated stationary
noise disturbances in each system, but also against systematic/spurious
disturbances that are not common to the two systems. The cross-correlation
spectrum \citep{Rubiola2010} is a well-known construct that allows
one to enhance a common signal -- here the hypothetical UDM signal -- in
two independent noisy detectors. Of course, one can attempt to minimize
correlated systematic disturbances by operating the two detectors
independently (e.g. without any common power supply) and sufficiently
distant from each other, so that they are subject to qualitatively
different environmental disturbances, but this is in practice not
always feasible. If independent operation is implemented, it will
be beneficial also when the cross-correlation analysis is applied.
\clearpage{}

\section{Concept of the apparatus}

In each detector, we measure the difference $\nu(t)$ between the
frequency $\text{\ensuremath{\nu_{L}}}$ of a laser and of a molecular
transition, $\nu^{(2)}$, as a function of time. Since the molecular
transition is interrogated by the laser, necessarily the laser frequency
$\nu_{L}$ needs to be chosen to be within the spectral width of the
molecular transition line. In practice, the laser frequency is stabilized
to an evacuated, rigid optical cavity (a ``reference'' cavity),
with resonance frequency $\nu^{(1)}$, in order to achieve a high
sensitivity also for small oscillation frequencies $f_{\phi}$ of the UDM field.
Thus, $\nu_{L}=\nu^{(1)}$ is ensured by a feedback system.

The dependence (sensitivity) of the frequency $\nu^{(i)}$ of any physical system
$i$ on a particular FC $g$ is given by the fractional derivative
$R_{g}^{(i)}=\partial\,{\rm ln}\,\nu^{(i)}/\partial\,\,g$. If any
FC $g$ oscillates with amplitude $\delta g$, this results in a modulation
of the detuning $\nu=\nu^{(1)}-\nu^{(2)}$ with an amplitude $\delta\nu/\nu=\Delta R_{g}\times\delta g/g$,
the net sensitivity being $\Delta R_{g}=R_{g}^{(1)}-R_{g}^{(2)}$.

It is desirable to employ a molecular species and a transition whose
sensitivity $R_{N}^{(2)}$ to the mass of the nuclei is as large as
possible. This sensitivity stems almost completely from the difference
in vibrational energy of the lower and upper state involved in the
transition. Each energy contribution is, in the Born-Oppenheimer approximation,
given by $E_{{\rm vib}}\simeq h\,c\,R_{\infty}(v+1/2)\gamma\sqrt{m_{e}/M_{N}}$.
Here, $v$ is the vibrational quantum number of the state, $\gamma$
is a molecule-specific, electronic-state-specific numerical constant, and $M_{N}$
is an effective (reduced) nuclear mass. The rotational energies contribute
much less to the mass sensitivity. Electronic transitions in molecules
have vibrational contributions, but in ``standard'' molecules, such
as iodine (I$_{2}$), the vibrational contribution is modest relative
to the purely electronic contributions. In a previous experiment \citep{Oswald2022},
two electronic transitions in I$_{2}$ at 532 and 725~nm were
employed. They exhibited sensitivities $R_{N}\simeq-0.06,\,0.07$,
respectively. Here, we use one particular component of the $\nu_{1}+\nu_{3}$
combination vibrational band of $^{12}{\rm C}_{2}{\rm H}_{2}$ \citep{Gilbert2001},
specifically line P(17) at 1535.4~nm. For the neighboring component
P(16), Constantin \citep{Constantin2016a} has computed $R_{N}^{(2)}=-0.468$,
very close to the Born-Oppenheimer value of $-1/2$. This value is
a good approximation for our line P(17) as well, implying an approximately
sevenfold improvement compared to the previously used iodine line
at 532~nm. The sensitivity to the electron mass and to the fine-structure
constant are $R_{\alpha}^{(2)}\simeq2$ , $R_{m_{e}}^{(2)}\simeq1-R_{N}^{(2)}$,
where the contributions ``2'' and ``1'' stem from the proportionality
of any molecular transition frequency to the Rydberg constant.

For the evacuated optical cavity, the frequency has the well-known
dependence $\nu^{(1)}\propto L^{-1}\propto a_{0}^{-1}=m_{e}\alpha\,c/\hbar$,
where $L$ is the mirror distance, and $a_{0}$ is the Bohr radius.
Therefore, the sensitivities of the cavity are $R_{\alpha}^{(1)}=1$,
$R_{m_{e}}^{(1)}=1$, $R_{N}^{(1)}=0$.

In total, the measured frequency difference $\nu$ has the combined
sensitivities\vspace{0cm}

\begin{align}
\frac{\delta\nu}{\nu} & =\Delta R_{\alpha}\frac{\delta\alpha}{\alpha}+\Delta R_{m_{e}}\frac{\delta m_{e}}{m_{e}}+\Delta R_{N}\frac{\delta m_{N}}{m_{N}},\quad\nonumber \\
 & =-\frac{\delta\alpha}{\alpha}-0.468\frac{\delta m_{e}}{m_{e}}+0.468\frac{\delta m_{N}}{m_{N}}.\quad\label{eq: delta nu/nu in terms of delta alpha/alpha, etc.}
\end{align}

Since any reduced nuclear mass $M_{N}$ is proportional to a generic
nuclear mass $m_{N}$, we have altogether from Eqs.~(1)--\ref{eq: m_N(phi)})
\vspace{0cm}

\begin{align}
\frac{\delta\nu}{\nu} & =\left(\Delta R_{\alpha}d_{\alpha}+\Delta R_{m_{e}}d_{m_{e}}+\Delta R_{N}(Q_{N}\cdot d_{N})\right)\,\phi.\quad\label{eq: delta nu/nu =00003D sum (Delta R d) x delta phi}
\end{align}

\vspace{0cm}
For comparison, our previous ``experiment A'' in Ref.~\citep{Oswald2022}
had sensitivities\vspace{0cm}
\begin{equation}
\frac{\delta\nu}{\nu}=-\frac{\delta\alpha}{\alpha}-0.06\frac{\delta m_{e}}{m_{e}}+0.06\frac{\delta m_{N}}{m_{N}}.\quad\label{eq:delta nu/nu of experiment A in Oswald et al}
\end{equation}

\vspace{0cm}
\clearpage{}

\section{Experimental setup}

\begin{figure}[H]
\centering{}\includegraphics[width=1\textwidth]{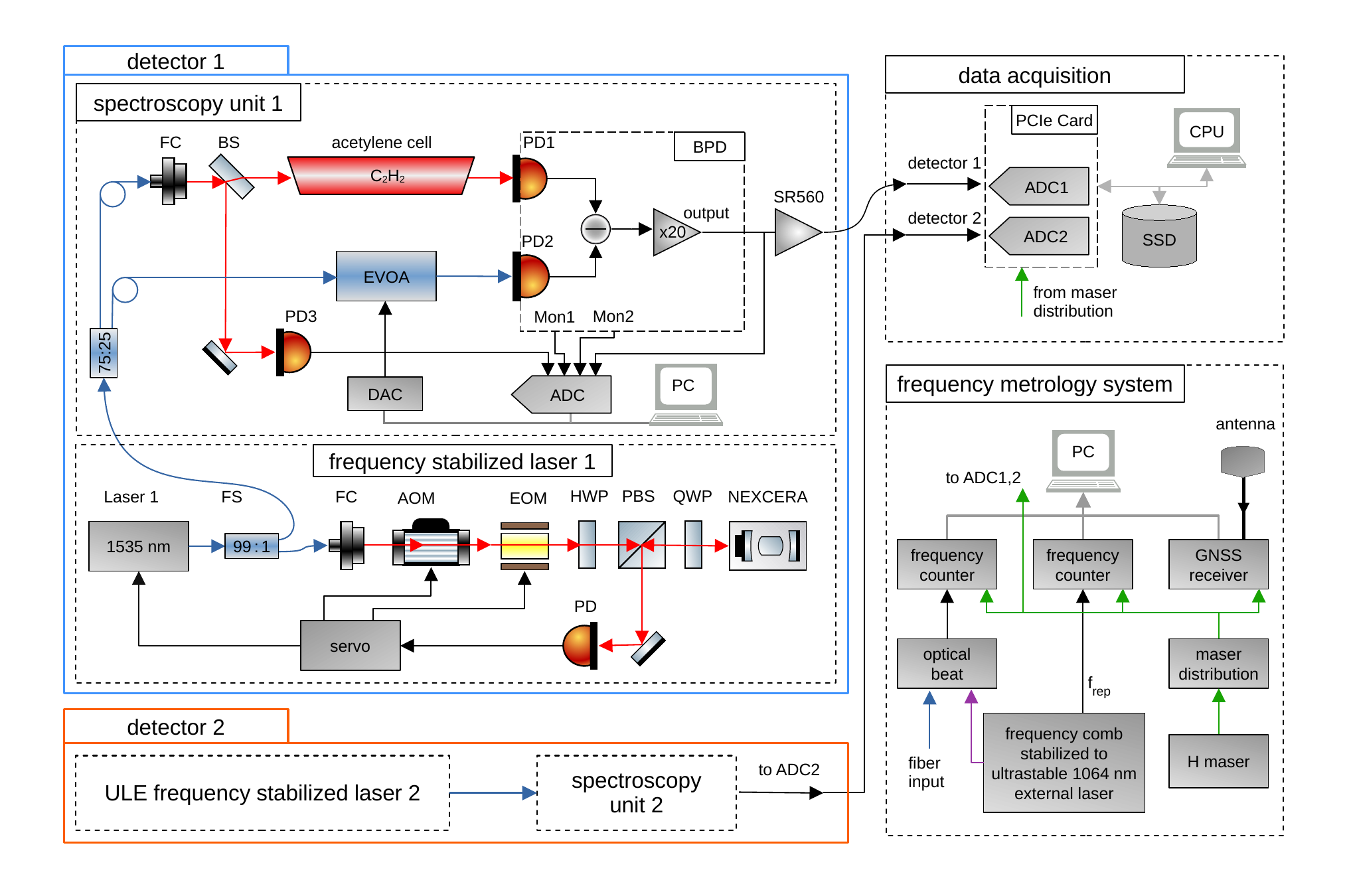}\caption{\label{fig:setup}Schematic of the paired UDM detector. Dashed
boxes show subsystems. \textbf{\textcolor{blue}{Blue}} lines, fiber-optical
connections; \textbf{\textcolor{red}{red}} line, free-space laser
beams; \textbf{\textcolor{purple}{purple}} line, comb radiation at
1.5~$\mu$m; \textbf{\textcolor{black}{black}} lines, analog electrical
connections; \textbf{\textcolor{gray}{grey}} lines, digital connections;
\textbf{\textcolor{green}{green}} lines, 10~MHz reference signal;
FC, fiber outcoupler; FS, fiber splitter; BS, beam sampler; EVOA,
electronic variable-optical attenuator; PD, photodetector, BPD, balanced
photodetector; SR560, low-noise amplifier and filter; AOM, acousto-optical
modulator; EOM, electro-optical modulator; HWP, half-wave plate; QWP,
quarter-wave plate; PBS, polarizing beam splitter; NEXCERA, ultra-low
thermal expansion ceramic; ULE: ultra-low-expansion glass. SSD, solid-state
disk; GNSS, global navigation satellite system.}
\end{figure}

A scheme of the apparatus configuration is presented in Fig.~\ref{fig:setup}.
The experiment uses two independent and nearly identical UDM detectors,
here called detector 1 and detector~2. They were installed in different
laboratories located approximately 10~m apart. Detector 1 received
the laser light from laser system~1 via a 5-m-long polarization-maintaining
polarization-maintaining fiber. Detector~2 was connected to laser system 2 via a 50-m-long
single-mode fiber. The two lasers were housed in the same laboratory
as detector~2.

Each detector contains one oscillator, a laser stabilized to a high-finesse
reference resonator ($\nu^{(1)})$, and a molecular-gas vibrational
spectroscopy unit. The molecular resonance frequency defines $\nu^{(2)}$.

For technical details we refer to the Supplemental Material.
Briefly, in each system, the stabilized laser frequency is tuned to
approximately the half-width of the Doppler-broadened molecular transition.
Any variation of the frequency difference $\nu=\nu^{(1)}-\nu^{(2)}$
between laser frequency and molecular resonance frequency translates
into a variation of the laser power transmitted through the molecular-gas
cell and therefore a variation of the differential photocurrent
signal from a balanced photodetector (BPD). This signal is acquired
with one channel of an analog-to-digital converter (ADC). The second
detector outputs its signal to the second channel of the same ADC,
thus ensuring that the two signals are acquired perfectly synchronously.
In addition, the ADC clock is referenced to a hydrogen maser. We acquire
data from both detectors during uninterrupted runs, each lasting 19
hours. The photocurrent signal is converted into frequency deviation
units (Hz) using the measured amplifier gains and the slope of the
molecular line shape. The time interval between the end of one run
and the following varies.\clearpage{}

\section{Experimental protocol and data processing\label{sec:protocol}}

\subsubsection{Data acquisition}

During each run, indexed by $m=1,\ldots,14$, the outputs of the two
BPDs were amplified, separately bandpass-filtered (SR560) and then
synchronously sampled by two ADCs inside a two-channel PCIe card at
244~140 samples per second and the data streamed to memory. The time base
of the card was connected to the 10~MHz reference output from a hydrogen
maser to ensure long-term stability. The two data recordings were
subsequently separately fast-Fourier-transformed and rescaled to provide
the lists ${\rm FFT}_{j,m}$, the Fourier spectrum of the variations
$\delta\nu$ of the cavity-molecular-vibration frequency difference $\nu^{(1)}-\nu^{(2)}$
of each detector $j=1,2$. The unit of the list elements is Hz.

All runs were vetted by inspection of the aggregated time-series
and frequency-domain data to ensure that there were no obvious excessive
signal excursions. The term ``aggregation'' is explained below.

\subsubsection{Calculation of the cross-spectrum}

In order to achieve a suppression of non-common-mode noise, we apply
the technique of cross-spectrum analysis \citep{Rubiola2010}. It relies on the possibility
of averaging a large number of cross-spectra taken independently over
time. Stationary noise that is uncorrelated between two detectors
is thereby averaged down. In addition, spurious signals appearing
independently in the detectors are also suppressed.

First, we compute $S_{12,m}$, the cross-spectrum of each run $m$,
from the fast-Fourier transforms (FFTs) of detectors of detector 1 and 2, ${\rm FFT}_{1,m}$ and ${\rm FFT}_{2,m}$,
as

\vspace{0cm}
\begin{equation}
S_{12,m}={\rm Re}\left\{ {\rm FFT}_{1,m}\times{\rm FFT}_{2,m}^{*}\right\}.\label{eq:CC}
\end{equation}

\vspace{0cm}
Then, the averaged cross-spectrum over $M$ experiments is the mean
value,

\vspace{0cm}
\begin{equation}
S_{12}=\frac{1}{M}\sum_{m=1}^{M}S_{12,m}.\label{eq:ACS}
\end{equation}

\vspace{0cm}
$S_{12}$ has unit ${\rm Hz^{2}}$.\\

\section{Determination of the detection limits}

In order to determine the upper bounds for the UDM coupling constants
$d_{g}$ we found it useful to apply synthetic signals to the data.
We now explain the procedure.

\subsubsection{Definition of the synthetic signal}

We numerically create a synthetic spectrum ${\rm FFT}(T)$ that emulates
the effect of an UDM spectrum. ${\rm FFT}(T)$ can be added to the
experimental spectra of the various runs. Here, $T$ is a time
domain representation related to a synthetic UDM field $\phi(t)$,
impacting an observable or a fundamental constant $g$ according
to  $\delta g(t)=d_{g}\,\phi(t)\times g$ and is defined further below. $\phi(t)$ is the dimensionless
UDM field amplitude for a particle of Compton frequency $f_{\phi}$
\citep{Centers2020,Derevianko2018,Oswald2022},

\vspace{0cm}
\begin{equation}
\phi(t)=\frac{\phi_{0}}{2\pi f_{\phi}}\sum_{i}\alpha_{i}F'(f_{i})\,{\rm cos}(2\pi f_{i}\,t+\varphi_{i}).\label{eq:phit}
\end{equation}
$\phi_{0}=\sqrt{4\pi\rho_{DM}G_{N}/c^{2}}$ $\simeq7\times10^{-16}$~Hz
is the normalized field amplitude of the UDM oscillation, $\rho_{DM}\simeq0.3~{\rm GeV/cm^{3}}$
is the local DM density, $G_{N}$ is the gravitational constant and
$c$ is the speed of light. The dimensionless amplitudes $\alpha_{i}$
are taken from a Rayleigh distribution with $\sigma_{\alpha_{i}}=1$.
The phases $\varphi_{i}$ are taken from a uniform distribution over
the range of $0-2\pi$, while $\{f_{i}\}$ is a set of frequencies
sampling the distribution $f_{DM}$, spaced by $\Delta f$. The dimensionless
$F'(f_{i})$ is defined as $F'(f_{i})=\sqrt{f_{DM}(f_{i})\Delta f}$,
where $f_{DM}(f)$ is a dimensional function that describes the UDM
line shape around the Compton frequency $f_{\phi}$. The function
is reproduced here from Eq.~(3.35) of \citep{Savalle2020Thesis},

\vspace{0cm}

\begin{equation}
f_{DM}(f_{i})=\frac{2}{\sqrt{2\pi}f_{\phi}v_{g}v_{vir}}{\rm exp}\left(\frac{2\left(\frac{f_{i}}{f_{\phi}}-1\right)+v_{g}^{2}}{2\,v_{vir}^{2}}\right){\rm sinh}\left(\frac{v_{g}}{v_{vir}^{2}}\sqrt{2\left(\frac{f_{i}}{f_{\phi}}-1\right)}\right).\label{eq:fdm}
\end{equation}

\vspace{0cm}
Here $v_{g}\times c=220\,$km/s is the speed with which the Solar
System moves through the Galactic halo. The velocity spread of the
UDM particles is quantified by $v_{vir}\times c=150\,$km/s. The fractional
full width of $f_{DM}(f)$ at half maximum is approximately $\Delta f_{\phi}/f_{\phi}\simeq2\times v_{vir}^{-2}=5\times10^{-7}$.
The sum of $F'(f_{i})^{2}$ over all frequencies is equal to 1 \citep{Foster2018}.

We introduce the normalized and dimensionless function $T(t)$,

\vspace{0cm}

\begin{equation}
T(t)=\sum_{i}\alpha_{i}F'(2\pi f_{i})\,{\rm cos}(2\pi f_{i}\,t+\varphi_{i})\,\,.\label{eq:T(t)-1}
\end{equation}
\vspace{0cm}
$T(t)$ oscillates nearly harmonically between almost equal positive
and negative values, and the amplitude of this oscillation, of order
$1$, varies weakly on timescales much smaller than $(f_{\phi}v_{vir}^{2})^{-1}$.
On this and longer timescales, however, the amplitude varies substantially,
even approaching zero at times \citep{Centers2020}. The long-time
mean of $T(t)^{2}$ is equal to 1 (effectively, Parseval's theorem).

The UDM field is then expressed as\vspace{0cm}

\begin{equation}
\phi(t)=\frac{\phi_{0}}{2\pi f_{\phi}}T(t).\label{eq: phi(t) =00003D ... T(t)}
\end{equation}

\vspace{0cm}
Since expression (\ref{eq:T(t)-1}) is already a Fourier decomposition,
we can trivially generate the Fourier transform of $T(t)$, ${\rm FFT}(T)$:
We conveniently choose the frequencies $\{f_{i}\}$ to lie on the
same grid as the experimental FFT. Thus, $\Delta f$ introduced above
is set equal to the bin width of the experimental FFT. For practical
reasons, we limit the range of considered frequencies $\{f_{i}\}$
to $[f_{\phi},f_{\phi}(1+3\Delta f_{\phi}/f_{\phi})]$. We calculate
$F'(f_{i})$ according to Eq.~\ref{eq:fdm}. 

We generate a set of $m$ independent versions of a synthetic signal
FFT, each denoted by ${\rm FFT}(T_{m})$. These can be multiplied
by a factor $\beta$ (carrying the unit Hz) that describes the coupling
strength of the UDM field to the observable recorded by the detector.
These synthetic signals $\beta\,\times{\rm FFT}(T_{m})$ are then
added to the data of the corresponding experimental run $m$, and
the cross-spectrum is computed

\begin{equation}
S_{12,m}(\beta)={\rm Re}\left[({\rm FFT}_{1,m}+\beta\,{\rm FFT}(T_{m}))\times({\rm FFT}_{2,m}+\beta\,{\rm FFT}(T_{m}))^{*}\right].\label{eq:S12,m}
\end{equation}
\vspace{0cm}
 The use of different realizations $T_{m}$ is justified because
during our campaign the delay between the end of run $m$ and the
start of run $m+1$ was long and varied.

Finally, we compute the average of the cross-spectrum including the
synthetic UDM signal,\vspace{0cm}
\[
S_{12}(\beta)=M^{-1}\sum_{m}^{M}S_{12,m}(\beta)\,.
\]

\vspace{0cm}
Note that a UDM signal of the considered form generates, via the
terms $\beta^{2}\left|{\rm FFT}(T_{m})\right|^{2}$, a positive contribution
to $S_{12}$, irrespective of the sign of $\beta$ and therefore of
the sign of the coupling constant $d_{g}$. Figure~\ref{fig:Spectra-of-experimental data with added syntethic signals}~top
shows the effect of adding a synthetic signal with $f_{\phi}\simeq12\,$kHz
to real data. We recognize that a synthetic signal of strength $\beta=10$~Hz
yields a clearly visible signature. We furthermore display in the
same figure two examples meant to simulate spurious signals.
As can be seen in the middle and lower panels, there is no UDM signature
visible if the synthetic signal is added only to one detector's data
or different synthetic signals are added to each detector. These examples
are evidence that the cross-correlation is capable of suppressing
such spurious signals.\clearpage{}
\begin{center}
\begin{figure}[H]
\begin{centering}
\noindent\begin{minipage}[t]{1\columnwidth}%
\begin{center}
\includegraphics[width=0.5\textwidth]{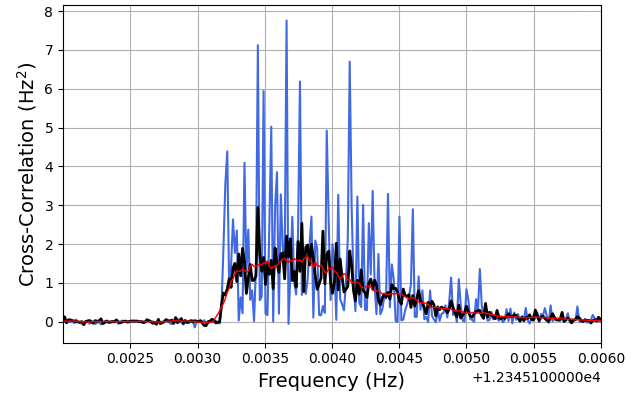}
\par\end{center}%
\end{minipage}\vspace{0bp}
\noindent\begin{minipage}[t]{1\columnwidth}%
\begin{center}
\includegraphics[width=0.5\textwidth]{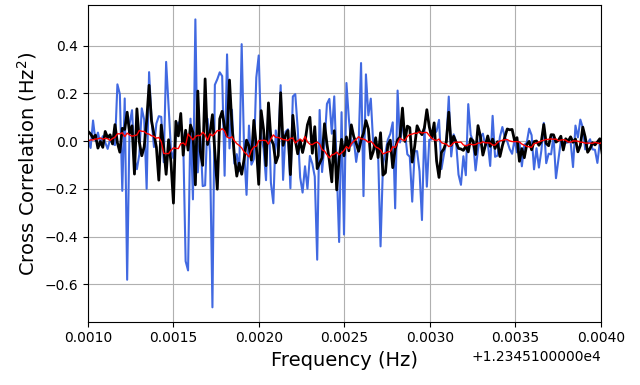}
\par\end{center}%
\end{minipage}
\par\end{centering}
\centering{}\vspace{0bp}
\noindent\begin{minipage}[t]{1\columnwidth}%
\begin{center}
\includegraphics[width=0.5\textwidth]{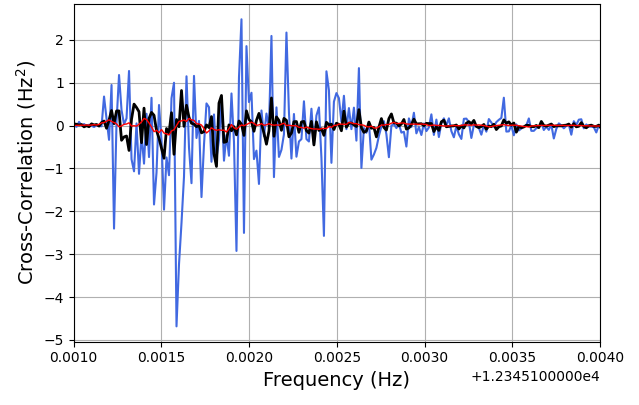}
\par\end{center}%
\end{minipage}\caption{\label{fig:Spectra-of-experimental data with added syntethic signals}\textbf{Spectra
of experimental data with added synthetic UDM signals. }Top panel:
in \textbf{\textcolor{blue}{blue}}, cross-spectrum $S_{12,m}(\beta=10\,{\rm Hz})$
of a particular experimental run $m$ with added synthetic UDM signal
($f_{\phi}=12.3451\,{\rm kHz}$) and in \textcolor{black}{black} the
average over 14 runs, $S_{12}(\beta=10\,{\rm Hz)}$.  Center and bottom
panels: hypothetical cases in which different synthetic signals (but with the same frequency $f_\phi$) are
added to the experimental data ${\rm FFT}{}_{1,m}$, ${\rm FFT}{}_{2,m}$.
Center panel: the synthetic signal is added only to ${\rm FFT}{}_{1,m}$
and the curves show ${\rm Re}\left[({\rm FFT}{}_{1,m}+\beta\,{\rm FFT}(T_{m}))\times({\rm FFT}{}_{2,m})^{*}\right]$
and its averages. Bottom panel: different realizations of the synthetic
signal are added to both ${\rm FFT}{}_{1,m}$ and ${\rm FFT}{}_{2,m}$.
The curves show ${\rm Re}\left[({\rm FFT}{}_{1,m}+\beta\,{\rm FFT}(T_{m}))\times({\rm FFT}{}_{2,m}+\beta\,{\rm FFT}(T_{m+1}))^{*}\right]$
and its averages. \textbf{\textcolor{blue}{Blue}} lines refer to
a particular individual experimental run $m$. \textbf{\textcolor{black}{Black}}
lines: averages over $M=14$ runs. \textbf{\textcolor{red}{Red}} lines:
0.124~mHz - running averages of $S_{12}$.}
\end{figure}
\par\end{center}

\subsubsection{Determination of the exclusion limits}

The exclusion limits are determined in several steps, based on the
experimental cross-spectrum $S_{12}(\beta=0\,{\rm Hz})$, the aggregated
cross-spectrum, denoted by aCS, and cross-spectra with added synthetic
UDM signals, $S_{12}(\beta\ne0)$.

\textbf{In step 1}, we aggregate 128 GiB data per detector (1 GiB $=2^30$ bytes) 
for each of the $M=14$ runs into one 64 GiB file. The duration of a run, 19
hours, implies that the FFT has a frequency resolution (bin width)
of 14 $\mu{\rm Hz}$. We compute the average $S_{12}(\beta=0\,{\rm Hz})$
from a total of 3.6~TiB of data. We then aggregate each group of 4096 subsequent
bins into one block having a width of 57~mHz, and generate the aggregated
cross-spectrum aCS. This is simply a list of sets, where each set $k$ 
is a summary of the cross-spectrum data $[S_{12}]_{k}$ in the $k$th
block that starts at frequency $f_{k}$:
minimum value (aCS-Min), maximum value (aCS-Max), mean value (aCS-Mean)
and standard deviation (aCS-SD),
\begin{align*}
{\rm aCS}(f_{k}) & =\{{\rm Min}( \{S_{12}\}_{k}),{\rm Max}(\{S_{12}\}_{k}),{\rm Mean}(\{S_{12}\}_{k}),\sigma(\{S_{12}\}_{k})\}.
\end{align*}

\begin{figure}[H]
\includegraphics[width=1\columnwidth]{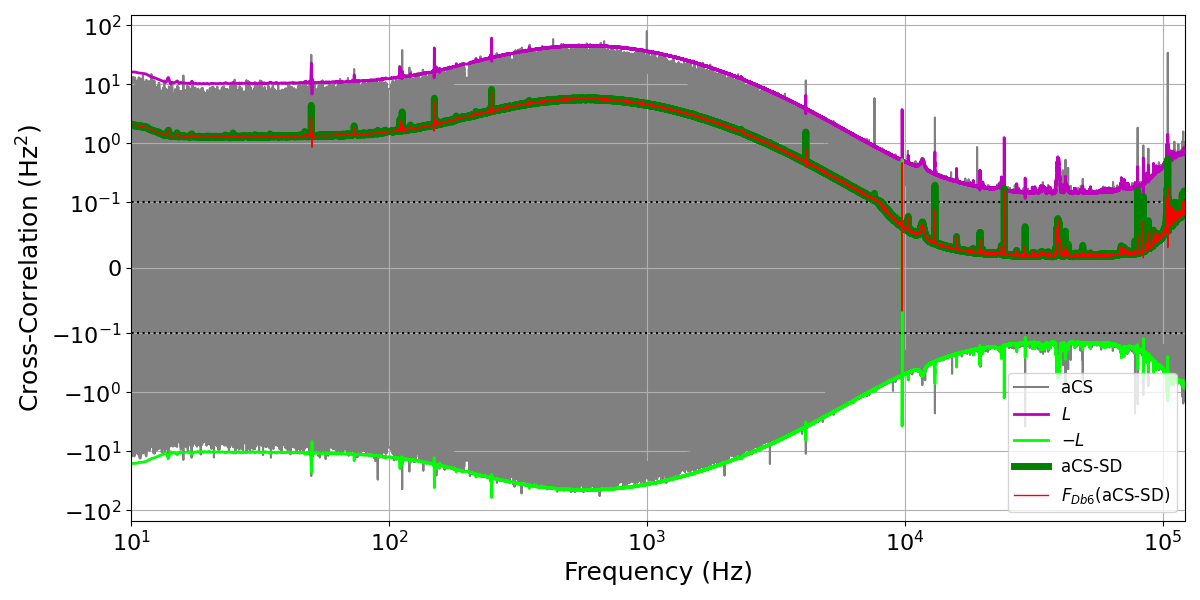}\caption{\label{fig:aggregated cross spectrum} \textbf{Properties of the aggregated
cross-spectrum aCS of $M=14$ experiments of 19-h-duration each.}
\textbf{\textcolor{gray}{Gray}}: minima and maxima of the aCS. Due
to the finite resolution of the graph, the gray data points are dominated
by the extreme values of the aCS-extrema. $2^{20}\simeq1\times10^{6}$
blocks are shown. \textbf{\textcolor{green}{Green}} line: aCS-standard
deviation (aCS-SD). \textbf{\textcolor{red}{Red}} line: Db6-filtered
version of the green line. \textbf{\textcolor{purple}{Purple}}: 8$\sigma$-bound
$L$ of the strength of a UDM signal. See text for details. Note
that the plot is composite: Above the black dotted line the vertical
scale is logarithmic, while it is linear below it. Technical noise
peaks were identified and removed from the spectrum.}
\end{figure}

The aCS is presented in Fig.~\ref{fig:aggregated cross spectrum}. The range between aCS-Max
and aCS-Min appears in \textbf{\textcolor{gray}{gray}}, the values
of aCS-SD are plotted in \textbf{\textcolor{green}{green}}. 

\textbf{In step 2}, a Daubechies sixth-order wavelet filter, denoted
as $F_{Db6}$, is applied to the aCS-SD so as to remove ``spikes''
that occur on a scale of one block. This procedure could remove UDM
candidates, but is nevertheless performed in order to generate a
UDM strength bound. This filtered aCS-SD is shown in \textbf{\textcolor{red}{red}}
in Fig.~\ref{fig:aggregated cross spectrum}.

We define the detection limit $L(f_{\phi})$ as the filtered \textbf{aCS-SD}
multiplied by a factor 8,
\begin{equation}
L(f_{k})=8F_{Db6}(\sigma(\{S_{12}\}_{k})).
\end{equation}
It is plotted as a magenta line in Fig.~\ref{fig:aggregated cross spectrum}.

\textbf{In step 3}, we compare $L$ with the unfiltered aCS-SD of
step 1 in order to identify outliers. Out of the $2^{20}$ blocks
we identified approximately 618 whose minimum lies below $-L$ and
approximately 583 whose maximum lies above $+L$ . Only the latter
blocks $\{k\}_{c}$ are inspected in step 5, because a UDM signal
would lead to a positive signal lying above $L$, as demonstrated
in Fig.~\ref{fig:Spectra-of-experimental data with added syntethic signals}.

\textbf{In step 4}, we determine the set of signal strengths $\beta^{*}(f_{k})$
required for the synthetic UDM signals $\beta^{*}(f_{k})\,T(f_{k})$
to cause a cross-spectrum value equal to $L(f_{k})$,
\begin{equation}
S_{12}(\beta^{*}(f_{k}))=L(f_{k})\,.\label{eq: condition for beta*}
\end{equation}
For a given frequency $f_{k}$ we added a synthetic signal of strength $\beta$
to the cross-spectrum and calculated the cross-correlation $S_{12}(\beta)$
as shown before in Fig.~\ref{fig:Spectra-of-experimental data with added syntethic signals}.
We applied a running average filter of $124\,{\rm \mu Hz}$ bandwidth
and determined the maximum of the filtered UDM signal. By adjusting
$\beta$ we matched this maximum to have the same amplitude as $L(f_{k})$
and store the final value as $\beta^{*}(f_{k})$.

We repeated this procedure not for all blocks $k$, but only for a
reduced number, 198, spanning the frequency range from 10~Hz to 122~kHz
and picked to be equidistant on a logarithmic frequency scale. The
frequency dependence of $\beta^{*}$ is presented in Fig.~\ref{fig:betastar}.
An example of a cross-spectrum containing a synthetic signal having
an amplitude $\beta$ close to the determined $\beta^{*}$ is shown
in Fig.~\ref{fig:candidate}. At the smallest analysis frequencies,
$f\simeq10\,{\rm Hz}$, the assumed physical DM signal falls entirely
into one bin only, and so does the synthetic signal. The strength
of the latter is then influenced by statistical sampling, and varies
strongly from one simulated realization to the next. However, because
we average over $M=14$ realizations, the variation averages out to
some extent, and our derived limits are reliable.
\begin{center}
\begin{figure}[H]
\begin{centering}
\includegraphics[width=0.8\textwidth]{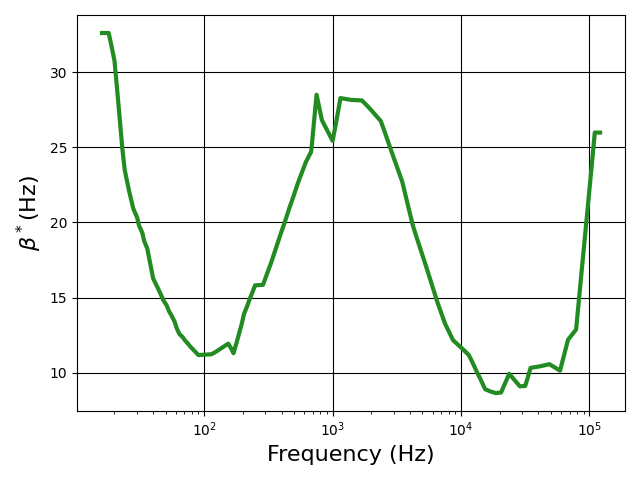}
\par\end{centering}
\caption{\label{fig:betastar}Determination of the minimum signal strength
$\beta^{*}$ of a synthetic signal necessary to be considered an UDM
candidate.  The \textbf{\textcolor{green}{green}} line connects the
set of 198 values $\beta^{*}(f_{k}).$ }
\end{figure}
\par\end{center}

\textbf{In step 5}, we inspect the blocks $k_{c}$ identified
earlier, where ${\rm Max}(\{S_{12}\}_{k})$ is above the limit $L(f_{k})$.
These blocks define candidate frequencies $\{f_{c}\}$. We then visually
inspected the experimental cross-spectrum $S_{12}(\beta=0)$ around
each frequency $f_{c}$ for any resemblance to a UDM line shape. An
example is presented in Fig.~\ref{fig:candidate}. To support the
inspection, we generated $S_{12}(\beta=\beta^{*}(f_{k}\approx f_{c}))$
and visually compared it with $S_{12}(\beta=0)$. Additionally, continuous
wavelet transformations (CWTs) of $S_{12}(\beta)$ and $S_{12}(\beta=0)$
are calculated. As the center panel shows, the CWT of a synthetic
signal creates a visually distinct pattern. Thus, CWTs help identify
possible candidates. Such a pattern is absent in the data; see bottom
panel. None of the 583 candidates could be confirmed as being consistent
with the assumed UDM line shape.
\begin{center}
\begin{figure}
\begin{centering}
\includegraphics[width=1\textwidth]{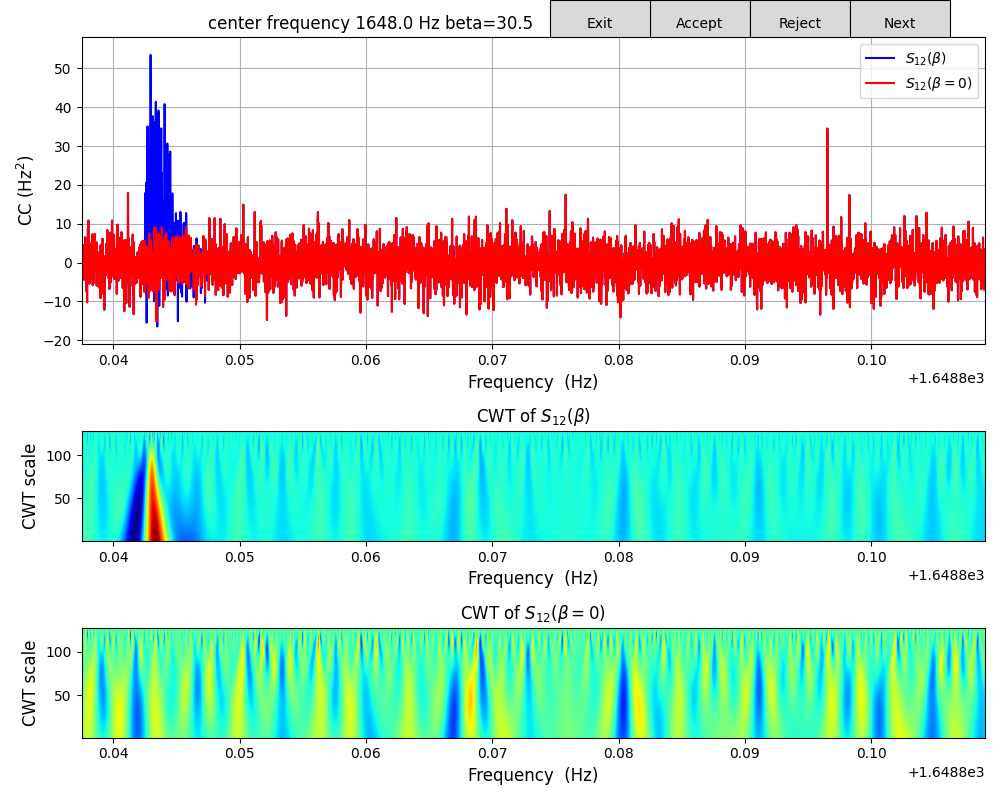}
\par\end{centering}
\caption{\label{fig:candidate}\textbf{Example for the visual inspection of
an UDM candidate} using a graphical user interface. Top panel: the
\textbf{\textcolor{red}{red}} spectrum shows $S_{12}(\beta=0)$. Note:
The red peak at approximately 0.095~Hz offset frequency is the candidate
that triggered the inspection. For comparison, the \textbf{\textcolor{blue}{blue}}
spectrum shows $S_{12}(\beta=30.5\,{\rm Hz)}$. Most of it is hidden
by the red trace. Center panel: CWT of $S_{12}(\beta=30.5\,{\rm Hz)}.$
Bottom panel: CWT of $S_{12}(\beta=0)$ CC: cross-correlation.}
\end{figure}
\par\end{center}

\textbf{In step 6}, we finally calculate the exclusion limit for a
coupling constant $d_{g}$ as such. As it is customary, we assume
in turn that only a single coupling constant is nonzero. We start
from Eq.~(\ref{eq: delta nu/nu =00003D sum (Delta R d) x delta phi}),
\[
d_{g}\phi=\delta g/g=(\Delta R_{g})^{-1}\delta\nu/\nu\,.
\]
The parameter $\beta$ was introduced to describe the effect of UDM
on the experimental observable, for the predetermined normalized time
dependence $T(t)$. Therefore we have the following equivalence, \vspace{0cm}
\[
\beta\leftrightarrow\frac{\phi_{0}}{2\pi f_{\phi}}d_{g}\times\Delta R_{g}\nu.
\]
\vspace{0cm}
The determined bounds $\beta^{*}(f_{k})$ therefore furnish the bounds
of the coupling constant, as\vspace{0cm}
\begin{equation}
d_{g}(f_{k})=\frac{2\pi f_{k}}{\phi_{0}}\frac{\beta^{*}(f_{k})}{\nu\,\Delta R_{g}}.\label{eq:bound for d_g}
\end{equation}
\vspace{0cm}
\clearpage{}

\section{\label{sec:Results-and-Discussion}Results and Discussion}

\begin{figure}
\begin{centering}
\noindent\begin{minipage}[t]{1\columnwidth}%
\includegraphics[width=1\textwidth]{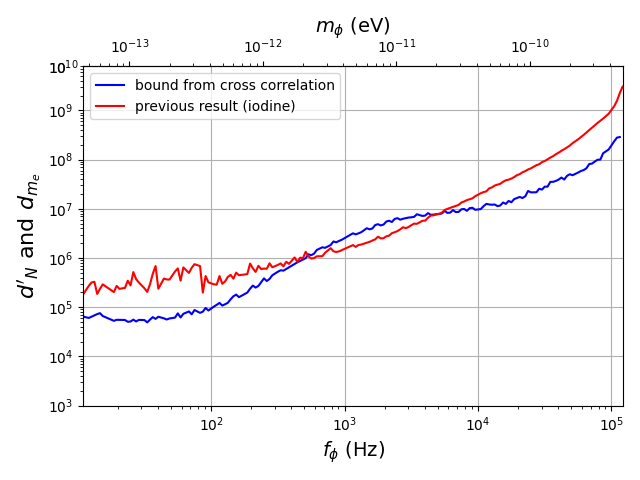} %
\end{minipage}
\par\end{centering}
\caption{\label{fig:dk}\textbf{Exclusion plot of the UDM coupling constant
to nuclear mass $d_{N}'$ and} \textbf{to electron mass} $d_{m_{e}}$.\textcolor{blue}{{}
}\textbf{\textcolor{blue}{Blue}} line: bound calculated from 272 hours
of measurement. \textbf{\textcolor{red}{Red}} line: bound from our
previous work on iodine spectroscopy, experiment A \citep{Oswald2022}.}
\end{figure}

Figure~\ref{fig:dk} shows the bound $d_{N}'$ obtained for the nuclear
mass. Here we define $d_{N}'=Q_{N}\cdot d_{N}$. The bound $d_{m_{e}}$
for the electron mass is the same, while it is approximately a factor
2 lower for $d_{\alpha}$; see Eq.~(\ref{eq: delta nu/nu in terms of delta alpha/alpha, etc.}).

In the range from 10 to 500~Hz we improve the bounds \textbf{$d_{N}'$}
and $d_{m_{e}}$ by a factor up to 10 compared to our previous iodine
experiment A \citep{Oswald2022}. In the range of 10 to 122~kHz we
improve by a factor up to 6. Our bounds for modulation of the nuclear
mass are currently the strongest ones, to the best of our knowledge.
Our bounds to $d_{\alpha}$ show only a small improvement by a factor
of 1.5 in the range below 80~Hz. This is related to the fact that
the shift from electronic molecular spectroscopy to vibrational spectroscopy
does not affect the sensitivity to $\alpha$; see Eqs.~(\ref{eq: delta nu/nu in terms of delta alpha/alpha, etc.}) and (\ref{eq:delta nu/nu of experiment A in Oswald et al}).

In conclusion, an important achievement of this work is the first
successful implementation of the cross-spectrum correlation technique
to UDM searches. We succeeded in developing efficient data handling
and analysis procedures. The number of spurious signals requiring
manual investigation and rejection was found to be much smaller than
in the iodine campaign. The impact in the low-frequency domain--below
100~Hz--has been quite beneficial. Note that the cross-spectrum
analysis technique is completely independent of the particular spectroscopy
approach used.

A second main result is proof that it is feasible to employ vibrational
spectroscopy for a UDM search. With the combination of the two techniques
introduced in this work, we achieve better bounds for two fundamental
constants than a previous iodine spectroscopy campaign, even though
in the latter the molecular transition linewidth was 160~times smaller
, as Doppler-free saturation spectroscopy was employed. The larger
linewidth in the present study implies lower sensitivity, but this
effect is partially offset by the stronger signal available in simple
absorption spectroscopy. In terms of technical simplicity and cost,
the acetylene experiment is advantageous: an iodine experiment requires
green laser light, for which the effort to provide a laser wave with
ultra-low frequency noise is higher than at the present wavelength
of 1.5~$\mu$m.

\subsection{Outlook}

This work may be extended in two ways. First toward higher Compton
frequencies, an extension that entails concomitant higher data volume
and processing effort if the frequency range is also extended. Second,
toward Doppler-free spectroscopy of acetylene \citep{Triches2015,Cherfan2020,Billotte2021},
promising lower noise and thus higher sensitivity for the spectral
window from tens of Hz to approximately 1~MHz. Looking towards the
future, the technique demonstrated here appears to be suitable to
be extended to a large number ($2N$) of simultaneously recording
detectors. Not every detector requires its own laser, since one laser
can serve as many detectors as its frequency noise level permits.
The lower the laser's frequency noise, the larger is the number. This
approach could ensure that the cost incurred in and volume occupied
by the large number of detectors $2N$ will not be prohibitive. The
most efficient way to process and make use of the $2N$ independent
data streams needs future investigations \citet{Derevianko2018}.
One possibility is to construct an $N$-pair cross-correlation $\Pi_{j=1}^{N}S_{2j-1,2j}$.
By construction, this quantity would suppress any oscillatory signal
that is not present in every detector.
\begin{acknowledgments}
We thank A.~Nevsky for performing exploratory work on acetylene
spectroscopy and E.~Wiens for the operation of the ULE-cavity-stabilized
laser system. U.~Rosowski and D.~Iwaschko helped with electronics.
This work has received funding from the DFG via project SCHI 431/24-1
(S.S.) and from both DFG and the state of Nordrhein-Westfalen via
Grant No. INST-208/774-1 FUGG (S.S.).
\clearpage{}
\end{acknowledgments}

\end{document}